\documentclass[10pt, final]{IEEEtran}

\usepackage{amsthm}

\usepackage{amssymb}
\usepackage{amsmath}
\usepackage{bbm}
\usepackage[ruled]{algorithm2e}
\usepackage{mathrsfs}
\usepackage{cite}
\usepackage{bm,epsfig,amsthm,url}
\usepackage{indentfirst}
\usepackage{amsfonts}
\usepackage{epstopdf}
\usepackage{cases}
\usepackage[caption=false,font=scriptsize]{subfig}
\usepackage{xcolor}
\usepackage{mathtools}

\makeatletter
\renewcommand{\maketag@@@}[1]{\hbox{\m@th\normalsize\normalfont#1}}%
\makeatother
\usepackage{hyperref}
\usepackage{stfloats}

\newtheoremstyle{mystyle}{}{}{}{}{}{: }{0pt}{\indent \it{\thmname{#1}\thmnumber{ #2}\thmnote{#3}}}
\theoremstyle{mystyle}

\usepackage{graphicx}
\usepackage{url}

\usepackage{tabularx,booktabs}
\newcolumntype{C}{>{\centering\arraybackslash}X} 
\usepackage{lipsum}

\usepackage{makecell} 

\usepackage{graphicx}
\usepackage{color}

\begin{document}
	

\title{ \fontsize{22pt}{26pt}\selectfont  Intelligent Reflecting Surfaces for Wireless Networks: Deployment Architectures, Key Solutions, and Field Trials}


	
\author{{Qingqing~Wu, Guangji Chen, Qiaoyan Peng, Wen Chen, Yifei Yuan, Zhenqiao Cheng, Jianwu Dou, \\Zhiyong Zhao, and Ping Li}
	\thanks{Q. Wu, Q. Peng and W. Chen are with Shanghai Jiao Tong University,  China ({qingqingwu, wenchen}@sjtu.edu.cn; qiaoyan.peng@connect.um.edu.mo). G. Chen is with Nanjing University of Science and Technology,  China (guangjichen@njust.edu.cn).  Y. Yuan is with  China Mobile Research Institute, China (yuanyifei@chinamobile.com). Z. Cheng is with China Telecom Research Institute, China (chengzq@chinatelecom.cn). J. Wu is with ZTE,  China ({dou.jianwu, zhao.zhiyong2, li.ping6}@zte.com.cn). 
	}
}

\maketitle

\begin{abstract}
Intelligent reflecting surfaces (IRSs) have emerged as a transformative technology for wireless networks by improving coverage, capacity, and energy efficiency through intelligent manipulation of wireless propagation environments. 
This paper provides a comprehensive study on the deployment and coordination of IRSs for wireless networks. By addressing both single- and multi-reflection IRS architectures, we examine their deployment strategies across diverse scenarios, including point-to-point, point-to-multipoint, and point-to-area setups. For the single-reflection case, we highlight the trade-offs between passive and active IRS architectures in terms of beamforming gain, coverage extension, and spatial multiplexing. For the multi-reflection case, we discuss practical strategies to optimize IRS deployment and element allocation, balancing cooperative beamforming gains and path loss. The paper further discusses practical challenges in IRS implementation, including environmental conditions, system compatibility, and hardware limitations. Numerical results and field tests validate the effectiveness of IRS-aided wireless networks and demonstrate their capacity and coverage improvements. Lastly, promising research directions, including movable IRSs, near-field deployments, and network-level optimization, are outlined to guide future investigations.

 
\end{abstract}


\vspace{-8pt}
\section{Introduction}
Intelligent Reflecting Surfaces (IRSs) have emerged as a cost-effective and transformative technology for next-generation wireless networks, offering promising applications and distinctive advantages \cite{wu2024PIEEE}.
Their potential has garnered substantial attention not only from academia, but also from industry leaders including Google, Apple, Qualcomm, Huawei, Samsung, and many mobile operators. In 2024, IRSs were recognized as one of the World Economic Forum's top 10 emerging technologies and prioritized in the U.S. national spectrum research and development plan as a key innovation area. Global initiatives, such as RISTA, IEEE ComSoc ETI, and efforts by standardization bodies like ETSI, IMT2030, and CCSA, further underscore the growing importance of IRS technology.

Current research on IRSs predominantly focuses on areas such as advanced signal processing (e.g., beamforming and channel estimation), the development of novel IRS architectures (e.g., passive and active designs), and their integration with other cutting-edge technologies, including massive/cell-free MIMO, mmWave, AI, and sensing/computing for applications across terrestrial, aerial, and maritime environments \cite{Basar2024VTM,multi-reflection_survey,wu2021intelligent}. However, research addressing IRS deployment strategies remains sparse, despite its critical importance for practical implementation.

From a network operator's perspective, optimizing IRS deployment is essential for achieving cost-effective and high-performance networks. Key considerations include IRS placement, size, architecture/type, densities, orientations, and reflection configurations. For instance, strategic IRS positioning can significantly enhance signal strength and coverage, especially in connectivity-challenged areas. 
Besides, properly deployed distributed IRSs are able to create rich-scattering environments, enabling multi-stream transmissions across the network and significantly boosting communication rates. Moreover, by employing multi-hop transmission with optimized reflective routing paths, robust line-of-sight (LoS) links can be established to blind spots to guarantee seamless coverage \cite{multi-reflection_survey}.

However, several critical challenges persist in the deployment and networking of IRSs. 
 First, the optimization of IRS deployment must be jointly coordinated with time, frequency, and spatial resource allocation to maximize overall network performance. For instance, different IRS deployment strategies may necessitate distinct multi-user access schemes, potentially leading to significant performance variations. 
 Second, the coexistence of heterogeneous network nodes—including multiple base stations (BSs), passive and active IRSs, and distributed user devices—coupled with diverse network topologies and complex node associations, poses substantial challenges for joint coordination and optimization. In particular, the mechanisms for achieving low-cost and energy-efficient networking with multiple IRSs remain underexplored and unclear. 
 Finally,  practical considerations such as power consumption, energy supply, environmental conditions, and maintenance requirements must be taken into account in deploying IRSs in commercial networks. These factors further compound the deployment of IRSs. 
 Addressing these challenges is paramount to unlocking the full potential of IRS technology, enabling better network performance, broader coverage, and cost-effective services.



In this paper, we present a systematic investigation into the deployment architectures and solutions of IRSs for various wireless network scenarios. Both single-reflection and multi-reflection IRS architectures are considered, with a focus on their deployment for point-to-point, point-to-multipoint, and point-to-area configurations. For single-reflection cases, we explore the trade-offs between passive and active IRS architectures, emphasizing beamforming gain, spatial multiplexing, and coverage extension. For multi-reflection cases, practical strategies are proposed to optimize IRS placement and element allocation, addressing challenges like cooperative beamforming gains and path loss. Simulation results and field trials validate the significant capacity and coverage improvements brought by IRS deployments. Furthermore, this paper highlights critical challenges, including hardware constraints and deployment complexity, while suggesting promising future research directions.


\begin{figure*}[ht]
    \vspace{-5pt}
	\centering
	\includegraphics[width=0.9\textwidth]{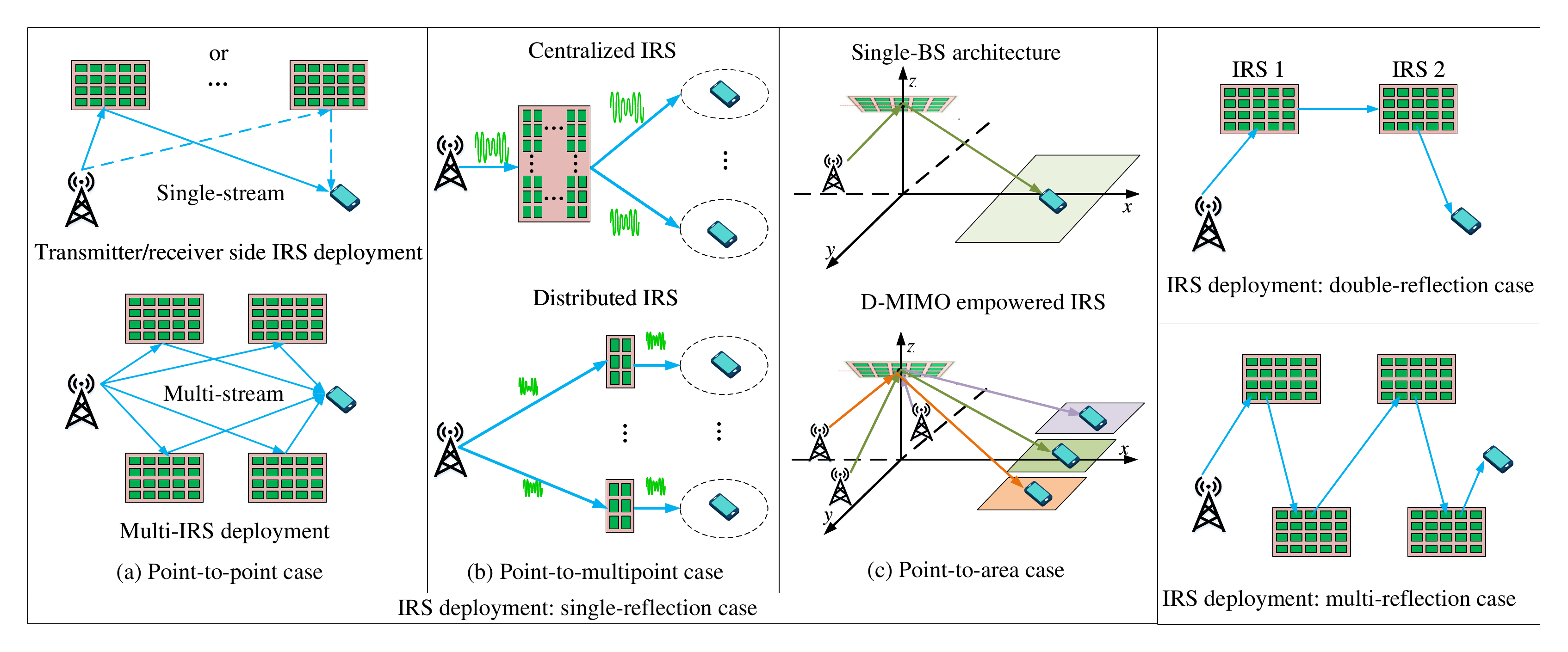}
	\vspace{-10pt}
	\caption{Illustration of representative IRS deployment scenarios.}
	\label{fig:systemmodel}
	\vspace{-8pt}
\end{figure*}

\section{IRS Deployment for Single-Reflection Case}

In this section, we provide a comprehensive overview of IRS deployment design issues in the single-reflection case.

\vspace{-4pt}
\subsection{Point-to-Point Setup}
This subsection examines an IRS-aided point-to-point setup, addressing key deployment design issues for both single- and multi-antenna cases. 


\subsubsection{Single-Antenna Case}
In the point-to-point case with a single antenna, the main objective of optimizing IRS deployment is to maximize the received power. For a passive IRS, this depends on the transmit power,  passive beamforming gain, and two-hop path-loss. Note that the placement of the IRS has a significant impact on the two-hop path-loss of the IRS reflecting link. To minimize this path-loss, a pioneering work \cite{wu2021intelligent} proved that a single IRS should be positioned either near the transmitter or receiver, resulting in two symmetric optimal placement options in the 2D plane, as illustrated in Fig. 1(a). 

For an active IRS with new signal amplifying capability, the received power of the IRS-aided link depends on the beamforming gain, two-hop path-loss, and the amplification factor. Hence, the deployment of active IRSs should balance the trade-off between two-hop path-loss and amplification factors. Given the power constraint of active IRSs, amplification factors are highly dependent on the first-hop channel. To maximize amplification factors and minimize two-hop path-loss, it is preferable to deploy the active IRS near the receiver \cite{R2}. However, this asymmetric deployment introduces a new challenge in balancing uplink and downlink communication quality. 
Notably, the SNR increase with the number of IRS elements is much slower for active IRS systems compared to passive ones. Inspired by this, a potential solution to address the aforementioned issue is to deploy two small-sized IRSs, one near the receiver and one near the transmitter, sharing the total number of elements. In a time-division duplex setup,  these IRSs are activated alternately in the time domain to assist the downlink and uplink communications, respectively. To this end, a theoretical framework is needed to shed light on how to deploy active IRS for balancing the communication qualities of both the uplink and downlink, which deserves further investigation.

Existing works demonstrated that passive and active IRSs possess complementary advantages. Motivated by their own benefits, a hybrid IRS architecture combining both the passive and active elements was proposed in \cite{R3} to achieve the fundamental trade-off between the unique power amplification gain of active IRS and a higher beamforming gain of passive IRS. Note that the deployment budgets (e.g., hardware cost and power consumption) for the active elements and passive elements are totally different. To balance the SNR-cost tradeoff, the number of active/pasive elements can be flexibly determined to achieve an energy-efficient architecture \cite{R4}. For another issue of hybrid IRS placement, a recent work \cite{R5} unveiled that the optimal position of a hybrid IRS is independent of the number of active/passive elements and its deployment principle is similar to the active IRS case for reaping the benefits of amplification.

\begin{figure}[t]
	\centering
    \vspace{-5pt}
	\includegraphics[width=2.5in]{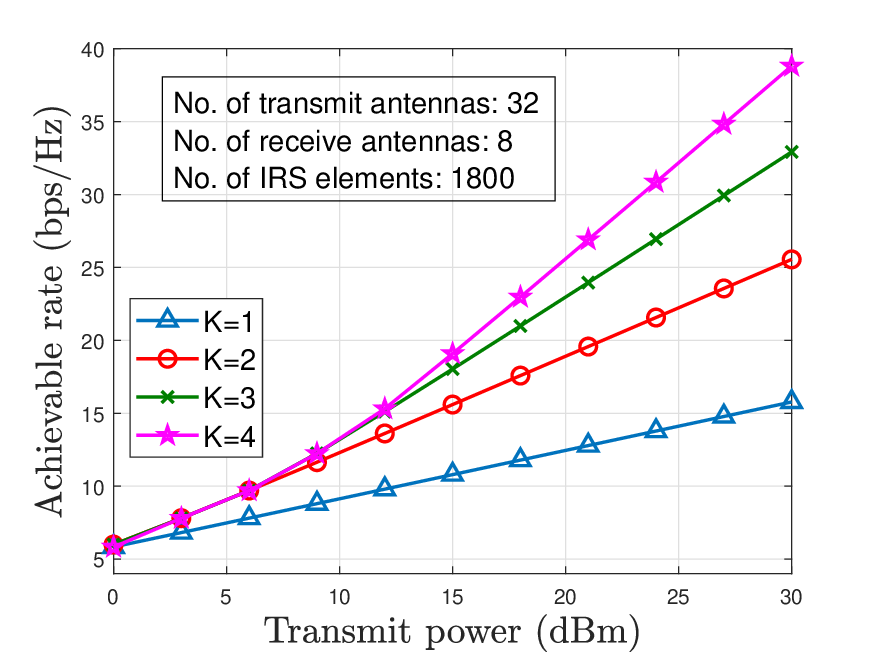}
    \vspace{-5pt}
	\caption{Capacity of a point-to-point MIMO assisted by $K$ IRSs.}
    \label{P2P_MIMO}
	\vspace{-10pt}
\end{figure} 
  
\subsubsection{Multi-Antenna Case}
With multiple antennas equipped at transceivers, multiple interference-free data streams can be transmitted simultaneously by exploiting the spatial domain. Note that the maximum number of data streams to be transmitted depends on the effective channel rank, meaning system capacity is influenced by both the received SNR and channel rank, which should be balanced in the IRS deployment design. In fact, properly deploying multiple IRSs can create a rich multi-path environment, improving channel rank and enabling parallel data stream transmission. Inspired by the potential of using multiple IRSs to address the rank deficiency issue, a novel spatial multiplexing-oriented channel reconfiguration approach was investigated in \cite{R6}. In particular, IRSs are placed in orthogonal directions of transceivers to create the orthogonal sub-channel. This effectively eliminates inter-stream interference, thereby unleashing the DoF of spatial multiplexing. The IRS element allocation is then optimized to balance the channel power gain for each data stream, which enables a flexible tradeoff between the beamforming gain and multiplexing gain.


To demonstrate the capacity improvement from multi-IRS deployment, Fig. \ref{P2P_MIMO} shows the system capacity versus transmit power for a point-to-point MIMO link assisted by multiple IRSs.  
It is observed that deploying multiple IRSs outperforms a single IRS, with gains becoming more significant as transmit power increases.  This highlights the advantage of using multiple small-size IRSs to increase the number of transmitted data streams in high-SNR regions. 
However, in low-SNR regions, the improvement is marginal as capacity is primarily limited by received power rather than channel rank.

\subsection{Point to Multi-Point Setup}
Next, we examine IRS deployment for multi-point scenarios with multiple users/clusters distributed far apart. Given a fixed number of IRS elements, two common deployment architectures are considered to minimize the two-hop path-loss: centralized IRS and distributed IRS. In the centralized architecture, all IRS elements are co-located near the BS. In contrast, the distributed architecture partitions all IRS elements into smaller IRS units, each deployed near a user cluster. The following discusses these two architectures for both single-antenna and multi-antenna cases.

\subsubsection{Single-Antenna Case}
In the single-antenna case, the effective channels from the BS to the $K$ users are reduced to $K$ scalars, making the capacity/achievable region of the IRS-aided multi-user system highly dependent on the corresponding channel power gains \cite{chen2022two,chen2022active,chen2022irs}. 
It was rigorously proven in \cite{R7} that a centralized IRS outperforms a distributed IRS in terms of the capacity region. This is because co-located IRS elements in a centralized architecture provide higher DoF to balance users' individual channel gains, while small distributed IRS units primarily enhance users within their local coverage. However, this advantage applies only to passive IRSs. For active IRSs, deployment must further account for amplification ability. Centrally deploying an active IRS near the BS significantly limits its amplification capability, suggesting that the optimal deployment architecture for active IRSs remains unclear and warrants further investigation.


\subsubsection{Multi-Antenna Case}
Equipping the BS with multiple antennas enables efficient spatial domain exploitation to serve multiple users simultaneously \cite{chen2023}. Consequently, IRS deployment must account for both user channel power gains and spatial multiplexing gains, differing fundamentally from the single-antenna case. A recent work \cite{R8} developed a theoretical framework to compare the system capacity of centralized IRS and distributed IRS deployment architecture under the homogeneous channel setup. By capturing their distinct channel characteristics, the study first identifies the capacity-achieving scheme for each architecture, which lays the foundation for performance comparison. 
For distributed IRSs, an ideal deployment condition can create orthogonal channels, allowing spatial division multiple access as the optimal scheme. In contrast, for centralized IRSs, highly correlated user channels which are common in LoS-dominated BS-IRS links necessitate a TDMA-based capacity-achieving scheme. Here, dynamic IRS beamforming maximizes each user’s channel gain during alternating transmissions.

\begin{figure}[t]
	\centering
    \vspace{-5pt}
	\includegraphics[width=2.5in]{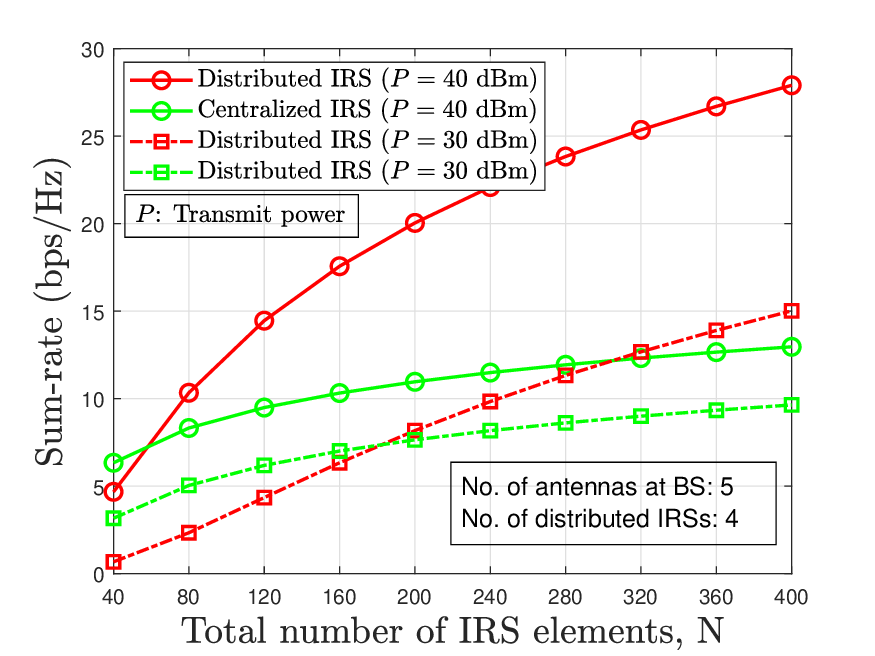}
    \vspace{-5pt}
	\caption{Sum-rates of two IRS deployment architectures.}
    \label{rate_N}
	\vspace{-10pt}
\end{figure}


To compare the capacity of the two IRS deployment architectures in a multi-antenna network, a numerical example for an IRS-aided broadcast channel is illustrated in Fig. \ref{rate_N}, showing the sum-rates versus the total number of IRS elements $N$. It is observed that the centralized IRS outperforms the distributed IRS when $N$ is small. In this low-$N$ region, system capacity is dominated by user-received power, and the centralized IRS benefits from a larger passive beamforming gain, significantly enhancing received power. However, as $N$ exceeds a threshold, the distributed IRS achieves higher sum-rates, with the performance gap widening as $N$ increases. This is because, in the large-$N$ region, the spatial multiplexing gain of distributed IRS is fully exploited. Hence, for practical systems with large $N$, the distributed IRS architecture is preferable for boosting the capacity of multi-antenna networks.

\subsection{Point-to-Area Setup}
In addition to boosting network capacity, IRSs can be effectively deployed to enhance wireless coverage in blind areas. 
Consider the setup in Fig. 1(c), where an IRS assists a BS in extending wireless coverage to a target area 
${\cal A}$. Here, the IRS placement and passive beamforming should be jointly optimized to maximize the minimum received power in ${\cal A}$. Specifically, the passive beamforming design must balance the array gain across the angular span associated with ${\cal A}$, with the objective value decreasing as the angular deviation increases, which depends on the IRS and target area's relative locations. Thus, IRS placement should balance angular deviation and two-hop path-loss. Prior work  \cite{R9} showed that BS-side IRS deployment minimizes both angular deviation and two-hop path-loss, leading to a higher beamforming gain. Therefore, BS-side IRS deployment is better suited for enhancing wireless coverage.


Unlike the  ${\cal O}\left( {{N^2}} \right)$ beamforming gain in point-to-point scenarios, only ${\cal O}\left( N \right)$ beamforming gain is achievable for point-to-area cases \cite{R9}.
 To address this limitation,  \cite{R10} proposed a novel IRS deployment architecture integrating the concept of distributed MIMO (D-MIMO) to enhance wireless performance over a target area. In this architecture, an IRS is deployed near a cluster of distributed BSs to extend coverage. By leveraging the spatial diversity of multiple BSs, the target area is divided into smaller sub-areas, each served by an associated BS. Through optimized BS-subarea associations, the angular deviation is significantly reduced, enhancing passive beamforming gain. Analytical results in \cite{R9} showed that with a sufficient number of BSs, the system can still achieve the 
 ${\cal O}\left( {{N^2}} \right)$ beamforming gain despite the static IRS phase-shifts.

\section{IRS Deployment for Multi-Reflection Case}

In complex environments (e.g., indoor-outdoor communication scenarios, unmanned factories, and millimeter and terahertz systems), double- or multi-reflection IRS are required, since their single-reflection counterparts may fail to bypass dense and scattered obstacles. Moreover, the passive beamforming gain of a single IRS is constrained by its physical size, with larger surfaces facing deployment challenges such as space limitations and maintenance. In contrast, multi-reflection IRS systems offer a viable alternative, enabling greater spatial diversity and higher cascaded beamforming gains. In double- or multi-reflection cases, channel state information can be obtained using estimation methods from existing works \cite{multi-reflection_survey}, e.g., the low-dimensional sub-channel decomposition technique and the joint estimation approach with always-ON training reflection.

\subsection{Double-Reflection Case}

This subsection first explores deployment strategies for two passive or active IRSs and then addresses design challenges for these architectures in double-reflection scenarios.

Similar to the single-IRS case, the optimal deployment strategy for a double passive IRS-aided system involves placing one IRS near the user and the other close to the BS. This setup enhances rate performance and coverage while maintaining signal quality and reliability. 
In a single-user scenario with double-reflection links only and LoS-dominant inter-IRS channels, a total of $N$ elements can achieve a passive beamforming gain of $\mathcal{O}(N^4)$ due to the multiplicative cooperative gain, with each IRS, equipped with $N/2$ reflective elements, contributing $\mathcal{O}(N^2)$ by aligning their beamforming directions \cite{doublePIRS_N}. Compared to the single-IRS case, achieving the passive beamforming gain of $\mathcal{O}(N^4)$ requires a large number of $N$ to compensate for the higher product-distance path loss. In the case with both double- and single-reflection links, a roughly equal element allocation strategy can be applied to maximize the rate by aligning the phase of the double-reflection link with the single-reflection link. Existing studies have shown that the power gain of the double-reflection link may fall below that of two single-reflection links when the number of IRS elements is insufficient and/or the inter-IRS channel is Rayleigh fading or NLoS. Therefore, the passive beamforming gains from the double- and single-reflection links should be effectively balanced according to the number of IRS elements and inter-IRS channel conditions.

Unlike double passive IRS-aided systems, deploying double active IRSs requires careful design to balance the multiplicative beamforming gain and amplification noise. The amplification power of active elements under different constraints significantly influences the deployment strategy and performance of double active IRS systems. For clarity, the IRS near the BS is referred to as IRS 1, and the one near the user is IRS 2. Previous studies have shown that under a total amplification power constraint, IRS 1 should be positioned farther from the BS as its number of IRS elements increases or its amplification power budget decreases. This allows IRS 1 to provide a larger amplification factor, compensating for cascaded path loss. From a power allocation perspective, more amplification power should be allocated to IRS 2 to mitigate noise with the increasing total number of active elements. 
In contrast, under a per-element amplification power constraint, IRS 1 should move closer to the BS to minimize noise amplification, while IRS 2 should be positioned directly above the user to reduce cascaded path loss as the number of IRS elements or per-element power increases \cite{doubleAIRS_xN}. In both cases, allocating more elements to IRS 2 is effective in reducing noise introduced by IRS 1 when the element budget permits.


Given the distinct deployment strategies for passive and active IRSs, how to effectively optimize their hybrid placement in a double IRS-aided wireless network is another key challenge. Depending on the order of the two IRSs, there are two transmission schemes: BS→active IRS→passive IRS→user (BAPU) and BS→passive IRS→active IRS→user (BPAU). In the BAPU scheme, the active IRS should be placed closer to the BS as the amplification power budget increases. This minimizes the cascaded channel gain and suppresses noise at the user. Conversely, under the BPAU scheme, the active IRS moves towards the user as the amplification power budget increases. It can provide a higher amplification factor and reduce cascaded path loss, resulting in a higher signal power received by the user. Another critical deployment design issue is determining element allocation between active and passive IRSs to enhance capacity.
Generally, more elements should be assigned to the active IRS when the amplification power budget (BAPU) or transmit power budget (BPAU) is sufficient, as it leads to higher amplification gain. In most cases, the allocation strategy favors deploying more passive elements as the total IRS element count increases. In most cases of the two schemes, the allocation strategy favors deploying more passive elements as the total number of IRS element increases. Practically, a joint design of element allocation and placement is required to further optimize system performance.

\begin{figure}[t]
	\centering
    \vspace{-5pt}
	\includegraphics[width=2.5in]{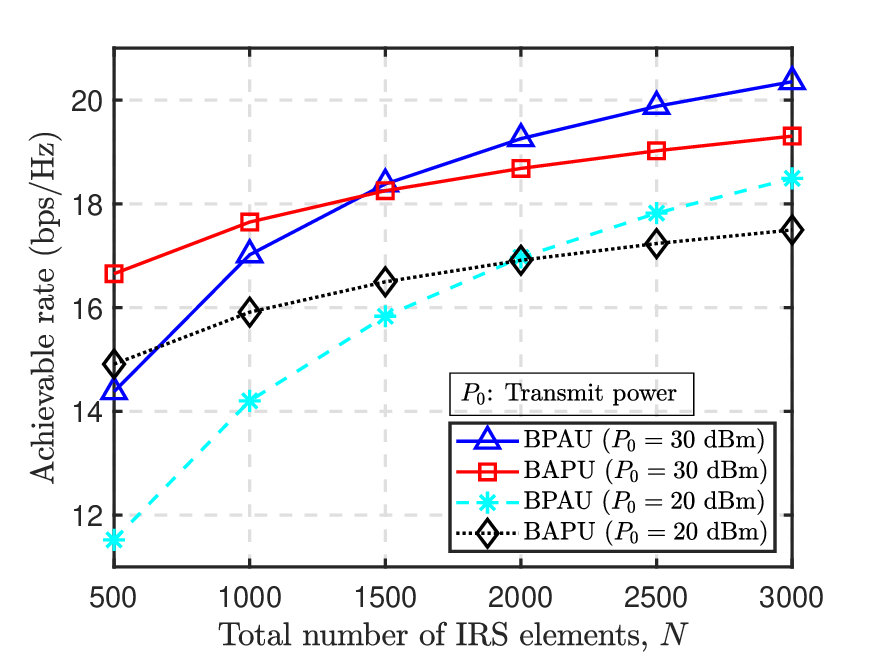}
    \vspace{-5pt}
	\caption{Capacity of two schemes with $P_0$.}
	\label{fig:R_opt_N}
	\vspace{-10pt}
\end{figure}

To illustrate the capacity of the BAPU and BPAU schemes with joint optimization, a numerical example of active and passive IRS-aided communication is presented. Fig. \ref{fig:R_opt_N} shows the achievable rate of the two transmission schemes versus the total number of IRS elements $N$. When $N$ is small, the BAPU scheme achieves a higher rate due to its amplification gain, allowing more power to reach IRS 2 and, subsequently, the user. However, as $N$ increases beyond a threshold, the BPAU scheme outperforms BAPU, as the passive beamforming gain of BPAU is fully realized in the high-$N$ regime.



\vspace{-5pt}
\subsection{Multi-Reflection Case}
In multi-reflection scenarios, particularly in densely deployed networks, IRS deployment presents significant challenges. These include optimizing factors not typically addressed in single or double IRS-aided systems, such as the number of IRSs, deployment costs, and interference among IRSs. This subsection begins with a discussion of multi-reflection scenarios involving passive IRSs and then extends to cases incorporating both passive and active IRSs.

From the perspective of network architecture, BS-side IRSs can save the high cost of densely deploying user-side IRSs and provide better overall system performance in the multi-reflection case with passive IRS. By jointly optimizing the BS and IRS deployment, the number of IRSs deployed or IRS deployment cost can be further decreased, without the need to increase the number of BSs. Existing works show that the coverage area gain increases proportionally with the ratio of the number of IRS elements to the BS-IRS distance \cite{multi-reflection_survey}. Moreover, leveraging intermediate users as relays can help reduce the deployment density of IRS in the environment, thereby preventing resource inefficiency. However, relying solely on intermediate users for IRS deployment may be not advisable as the unavailability of relays will result in unreliable delay-limited information transmission. In a large-scale network with massive randomly distributed users, several promising research directions worth exploring include developing dynamic deployment strategies and collaborative communication protocols to ensure reliable and efficient information transmission.

To further enhance the transmission coverage, the joint use of both active and passive IRSs may be considered. From the viewpoint of optimizing the performance in a single-cell setup with a single active IRS, existing works unveiled that all selected multiple-reflection links pass through the active IRS, which highlights the importance of carefully designing its location. The active IRS should be placed closer to the user as the number of passive elements increases. In a multi-cell network, incorporating active IRSs with optimal deployment design is beneficial for reducing the total deployment cost. In practice, candidate locations for IRS deployment can be predetermined via the ray-tracing approach based on the known topology of the region, which however incurs high deployment costs. To address this issue, the work \cite{x_multi} proposed increasing the number of non-overlapping cells with a selected subset of candidate locations. This is attributed to the additional DoF to optimize the IRS location in each cell but at the expense of a higher computational complexity of deployment optimization. Given the cell-use cost, the number of elements at each IRS should be optimized to balance between the hardware cost and system performance. As such, mathematical tools such as stochastic geometry and deep learning can be invoked to facilitate efficient element allocation. 
For the challenging case of multiple active/passive IRSs over a reflection path, considering the QoS requirements, location, user channel conditions, and compound channel fading due to multi-reflection propagation, it is worth studying AI-based IRS deployment solutions using e.g. federated learning and reinforcement learning.

\section{Practical Considerations for IRS Deployment}
\subsubsection{ {\textbf {Physical Deployment}}} 
First, the IRS placement must ensure effective coverage of the target area and efficient signal propagation. Despite having deployment guidelines in Sections II and III, obstacles like buildings and trees in practical site-specific environments can disrupt reflection paths, necessitating careful positioning of IRSs to avoid obstructions, consider dynamic changes in scatterers, and minimize unnecessary multi-path reflections to maximize efficiency. Moreover, potential gains can be achieved even if IRSs deployed by other operators cannot be controlled when an operator deploys IRSs, by strategically placing its own IRSs. To reduce interference, IRSs can be designed to operate within the allocated frequency bands, ensuring minimal leakage into other operators' bands. Second, power supply is another crucial consideration, especially for passive IRSs that require minimal but consistent energy for operation and active IRSs that demand higher power. Thus, deploying them in locations with readily accessible power sources helps avoid the cost and complexity of extensive power lines. In areas with limited power access, deploying novel IRS architectures, such as solar- or wireless-powered IRS, can enhance durability and achieve operational self-sustainability. 
Finally, IRSs are usually attached to existing structures such as walls, billboards, or rooftops, as free-space deployment is often impractical. As such, deployment locations and sizes are constrained by site rental costs and physical characteristics of available surfaces, including load-bearing capacity, space availability, and urban aesthetic considerations.

\subsubsection{{\textbf{IRS Practical Beam Pattern}}}

The non-ideal radiation pattern of IRS elements poses significant challenges for practical deployment by restricting the horizontal and vertical steering capabilities of the IRS beam to certain angular ranges. For example, it can lead to substantial variations in reflection channel gains for users located at different angles, thereby affecting the uniformity and efficiency of signal coverage. This limitation necessitates careful consideration of the orientation and placement of IRSs relative to BSs with practical antenna patterns and users \cite{ERP2}. On the other hand, the propagation environment, combined with multipath effects, further complicates IRS deployment as both the distance and angle between the IRS and the target receiver influence the quality of reflected signals. To enhance system performance, it is essential to accurately characterize the incident angle and configure the corresponding IRSs' reflections to effectively steer the beam towards target users. Moreover, system designers may consider the trade-offs between IRS element density and the required coverage area. Denser deployments can partially offset radiation pattern constraints but increase cost, complexity, and energy consumption.
	
\subsubsection{  {\textbf {System Integration and Scalability}}}
One of the main challenges in large-scale IRS deployment is effective scheduling and management, including network control and signaling overhead. Higher IRS densities mean more devices need to be integrated into the network, necessitating careful coordination to prevent interference and maintain performance. The operating frequency of IRSs must be compatible with existing systems to avoid interference. Moreover, more IRSs increase signaling overhead, leading to delays and reduced efficiency, especially in dynamic environments with mobile users. IRSs should be placed in stable channel conditions or where they can serve multiple users to reduce update frequency. Therefore, balancing IRS density with signaling costs is crucial. Deploying larger IRSs might reduce the corresponding number, providing a simpler, scalable solution. However, they require more power and control resources. Additionally, managing larger IRSs also increases signaling overhead due to the need for detailed information to adjust their reflection properties accurately.
	
\subsubsection{  {\textbf {Hardware Implementation and Maintenance}}}
a) Flatness: material loss, surface roughness, and manufacturing defects can lead to energy loss, affecting reflection efficiency. This variation in incident angle and frequency impacts the IRS's reflection configuration. 
b) Quantization Error: To maintain low hardware costs and power consumption, limited quantized phase shifts are adopted, leading to severe phase noise and affecting the accuracy of beamforming. 
c) Stability: IRSs must withstand environmental factors like wind, rain, and temperature fluctuations, as they operate outdoors for long periods. The electromagnetic response of an IRS is influenced by materials, manufacturing processes, temperature, and humidity, leading to unstable reflection characteristics and necessitating electromagnetic compatibility testing. 
d) Durability: Key components, including phase adjustment elements and control circuits, necessitate power, resulting in significant energy consumption. Efficient energy management is crucial to minimize power usage and extend the equipment's lifespan. 
e) Economic Considerations and Maintenance: The deployment and maintenance costs of IRSs, as well as performance enhancements, should be comprehensively evaluated to maximize economic benefit. Specifically, long-term maintenance is required, including fault detection and resolution. For example, a failure of the tuning elements at the IRS, including reduced beam gain, beam direction deviation, and increased side lobes.

    \vspace{-5pt}
\section{Field Tests}
\begin{figure}[!t]
\centering
\subfloat[Map of an IRS field test at 2.6 GHz.]{
		\includegraphics[scale=0.15]{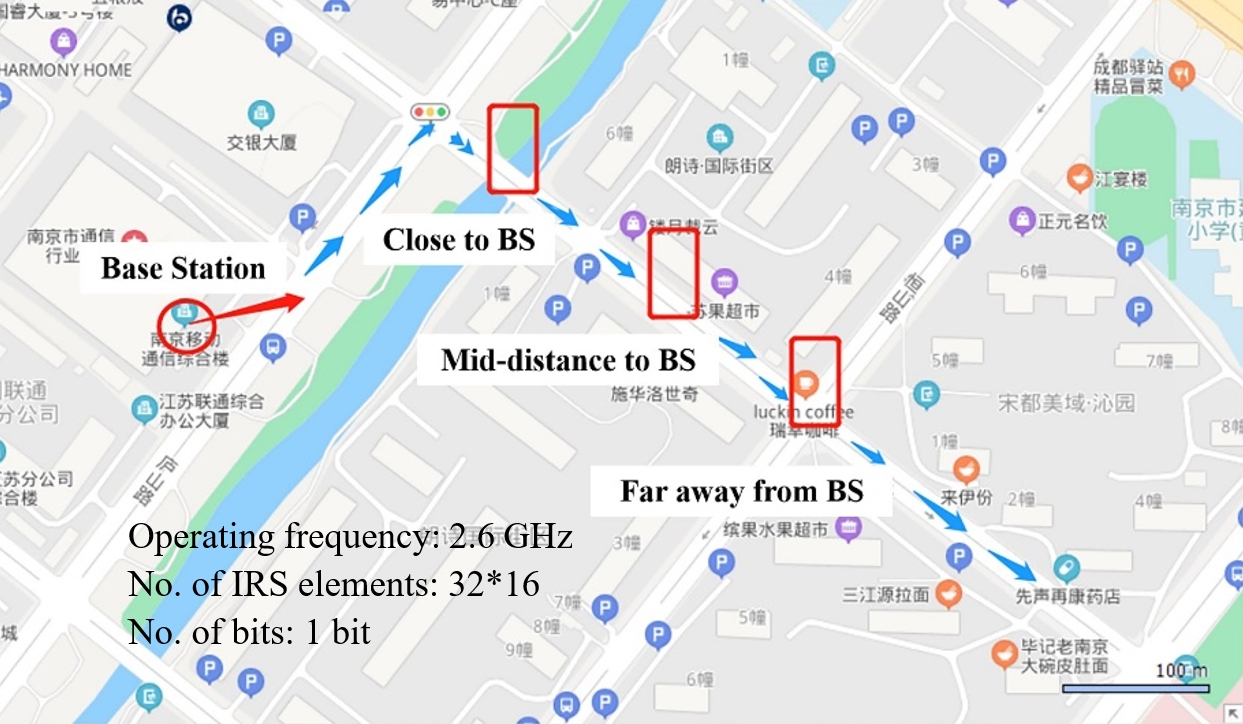}} \\
\subfloat[CDFs of RSRP.]{
		\includegraphics[scale=0.135]{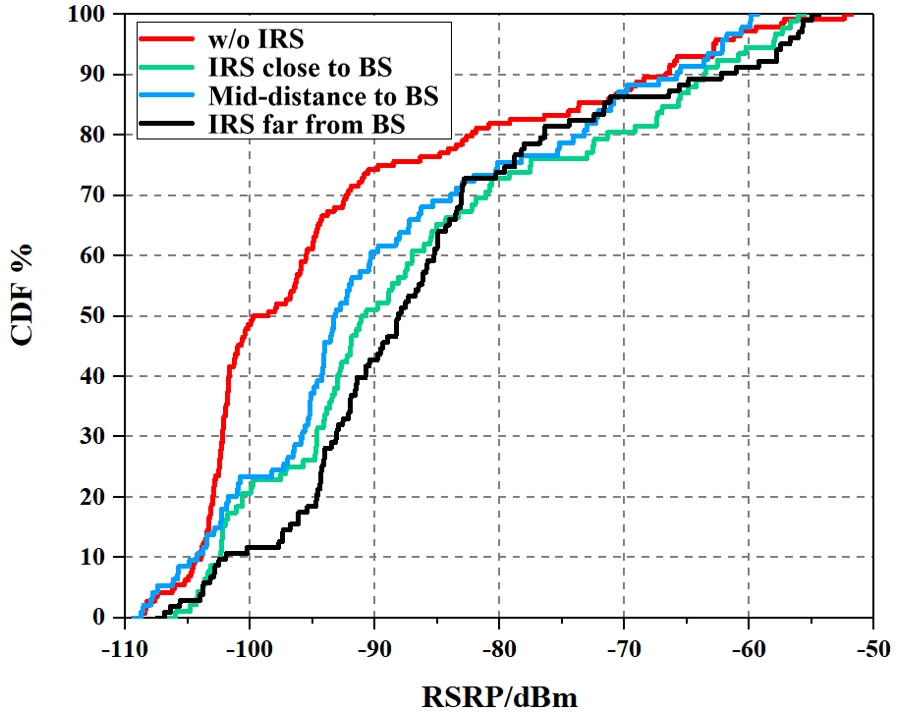}}
\hspace{-0.1cm}
\subfloat[CDFs of downlink throughput.]{
		\includegraphics[scale=0.16]{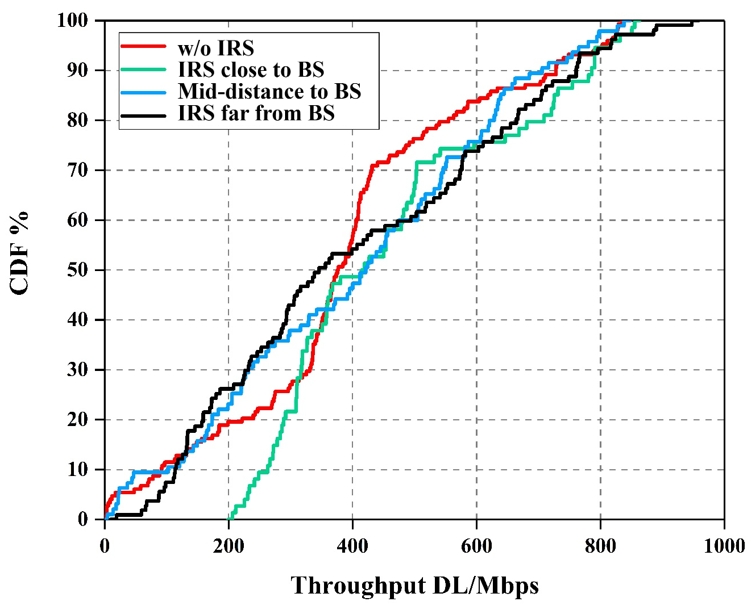}}
\caption{Field tests of deploying a single IRS in 5G commerical networks.}
\label{FT}
\vspace{-10pt}
\end{figure}

This section presents our two field trials of deploying IRSs in practical 5G commercial 
 and mmWave  networks. Fig. \ref{FT}(a) presents the layout of an IRS field test at the conducted in a typical urban area in Nanjing city. %
%
Three IRS locations are marked with red rectangles, labeled as ``Close to BS", ``Mid-distance to BS", and ``Far away from BS". 
At all three locations, the IRS panel is positioned on the ground beside the road, ensuring line-of-sight propagation between the BS and IRS. The IRS panel's orientation is optimized to balance the incident and reflected angles. As depicted by blue arrows, the test mobile moves along the roads, initially heading northeast, then turning right at the intersection to continue southeast. The phases of the IRS elements are adjusted to align the reflected beam along the southeast road direction, parallel to the ground.
Fig. \ref{FT}(b) presents the cumulative distribution functions (CDFs) of reference signal received power (RSRP) with and without IRS at the three tested locations. 

From Fig. \ref{FT}(b), it is evident that RSRP without IRS is suboptimal, with approximately 50\% of mobile locations recording RSRP below -100 dBm. The deployment of IRS significantly enhances RSRP, though the extent of improvement varies. The coverage improvement is least effective when the IRS is located at mid-distance to the BS, consistent with the analysis in Section II-A. When the IRS is positioned near the BS, the upper part of the CDF shows greater gains. Conversely, placing the IRS near the cell edges results in more substantial improvements at the lower part of the CDF. 
Fig. \ref{FT}(c) compares the CDFs of downlink throughput across the three IRS locations. Notably, when the IRS is positioned close to the BS, throughput gains are observed across various ranges. However, when the IRS is placed at mid-distance or farther from the BS, the performance gains are inconsistent, with mixed results. This phenomenon may be attributed to the use of massive MIMO  for downlink transmission. While massive MIMO operates dynamically, with spatial precoding updated every few milliseconds, the IRS in the field test is semi-static and effectively functions as a single antenna port. This limitation prevents the IRS from dynamically cooperating with the BS's massive MIMO, potentially reducing its overall performance impact.

\begin{figure}[!t]
\centering
\subfloat[Map of an IRS field test at 26 GHz.]{
		\includegraphics[scale=0.15]{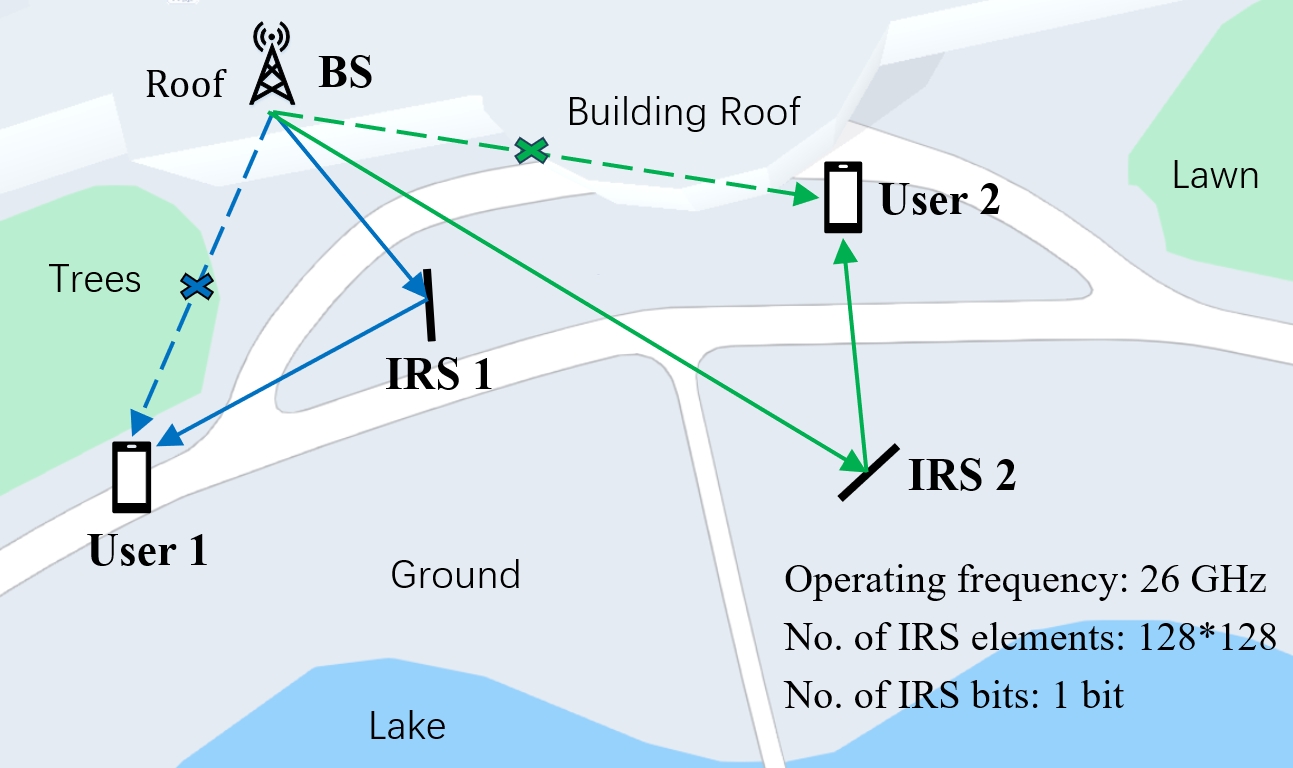}} \\
\vspace{-2pt}
\subfloat[Average RSRP.]{
		\includegraphics[scale=0.15]{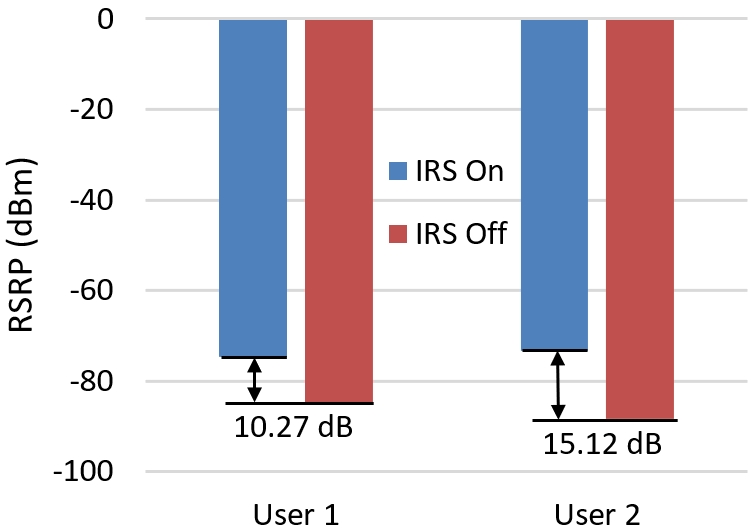}}
\subfloat[Average downlink throughput.]{
		\includegraphics[scale=0.15]{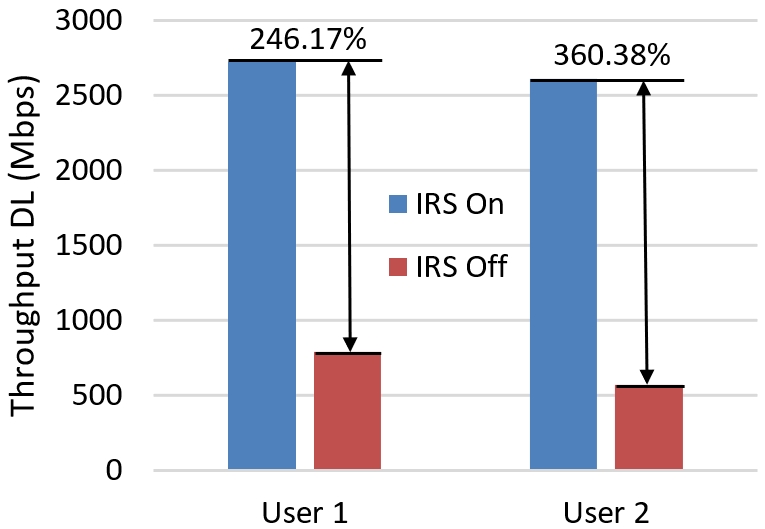}}
\caption{Field tests of deploying two IRSs in mmWave networks.}
\label{FT2}
\vspace{-10pt}
\end{figure}

Fig. \eqref{FT2}(a) presents the layout of a field test with two IRSs conducted in Shanghai city. It is observed from Fig. \eqref{FT2}(b)  that UE 1's RSRP increased from -84.98 dB (IRS off) to -74.71 dB (IRS on), an improvement of 10.27 dB, and UE 2's RSRP improved by 15.12 dB, from -88.39 dB to -73.27 dB. This is attributed to the IRS's ability to reflect and redirect wireless signals, thereby enhancing signal strength and coverage, reducing signal attenuation and interference, and ultimately increasing RSRP. Fig. \eqref{FT2}(c) compares the downlink throughput of the two users with and without IRS. It is evident that UE 1's downlink throughput increased by 246.17\% with IRS, from 785.94 Mbps to 2720.72 Mbps, and UE 2's downlink throughput increased from 566.34 Mbps (IRS off) to 2607.29 Mbps (IRS on), an increase of 360.38\%. This significant increase is due to the deployment of multiple IRSs with multiplexing-based beamforming and interference suppression techniques, which enables parallel transmission of multi-user data and achieves a substantial increase in total throughput. These results demonstrate that deploying IRSs can effectively boost signal strength and data transmission efficiency, thereby greatly improving the system's performance.

\section{Concusions and Future Directions}
This article offers a comprehensive overview of IRS deployment strategies in wireless networks, covering both single- and multi-reflection architectures, supported by numerical results and field tests. It aims to provide valuable insights and practical guidelines for deploying IRS-aided wireless networks. Despite recent progress, IRS deployment design still faces significant challenges that warrant further research. Several promising directions are outlined below. 

\subsubsection{  {\textbf {Movable/Rotatable IRS}}}
Recent works, such as \cite{R_a}, have explored movable-element IRS architectures to reconfigure channels dynamically. While prior works primarily focus on theoretical gains, practical challenges regarding the impact of mechanical latency, energy constraints for mobility, and real-time coordination on IRS deployment remain underexplored. We propose three architectures: 1) Movable antennas empowered fixed IRS: The positions of transceivers’ antennas are dynamically adjusted over time while the positions of IRS elements (with/without phase tunning) are fixed. 2) Fixed antennas empowered movable IRS: The positions of transceivers’ antennas remain fixed while the positions of IRS elements are adjustable. 3) Positions of both antennas and IRS elements are adjustable. Unlike existing efforts, our framework emphasizes \textit{dynamic environment adaptation} and \textit{scalable control mechanisms} to balance performance gains against mobility-induced overhead. Key directions include AI-driven position optimization, robustness against mechanical imperfections, and hybrid designs integrating movable subsurface with static IRS arrays for cost-effective deployment.

\subsubsection{  {\textbf {IRS Deployment for Multi-Function Network}}}
Future wireless networks are expected to support various new functions, including communication, sensing, computation, etc. For example, integrated sensing and communication (ISAC) aims to realize communication and sensing functions in unified systems. In particular, the objective of sensing is to extract target orientation information from the LoS channel, since the NLoS channel only contains interference or clutter \cite{SPS}. In contrast, NLoS channels are beneficial for being leveraged to enhance multiplexing gains. Despite recent works on optimizing IRS phase-shifts for ISAC, the deployment of IRSs can be designed to create artificial LoS/NLoS paths, which leads to a flexible performance tradeoff between communication and sensing.

\subsubsection{  {\textbf {IRS Deployment in Near-Field Region}}}
With the extremely high frequency and large-scale IRS/antenna array deployed in future communication systems, wireless transmission is expected to fall into the near-field region. The interplay between the features of near-field transmission and IRS deployment should be carefully captured to improve the performance of future wireless networks. Recall that the LoS channels exhibit a low rank in the far-field region. In contrast, by deploying a large IRS in the near-field region of a transmitter, the rank of the channel matrix of the transmitter-IRS link can be improved due to the nature of spherical wavefronts. By considering the enhanced spatial multiplexing ability in the centralized IRS architecture, the performance comparison between the centralized IRS and distributed IRS needs to be re-visited in the near-field region.

\subsubsection{  {\textbf {Networked IRS Deployment}}}
For a multi-cell network involving both single and multiple IRS reflection links, the densities of IRSs/BSs and number of elements/antennas at IRS/transceivers need to be jointly optimized for resolving the fundamental tradeoff between the network throughput and power consumption/cost. 
By capturing the topological randomness of BSs/IRSs, stochastic geometry has been identified as an advanced mathematical tool to characterize both the complicated inter-cell interference and intended signal induced by the reflections of IRSs. The initial stochastic geometry-based investigation \cite{Lyu2021} unveiled that the network throughput will be significantly improved by adding IRSs as compared to the network with BSs only. However, the work \cite{Lyu2021} is limited to the case of single-antenna at the BS. By incorporating new network architectures or IRS architectures, e.g., IRS-aided cell-free MIMO and active IRS, the network capacity limits remain largely unknown, which is a promising research direction.
\vspace{-5pt}


\bibliographystyle{IEEEtran}
\bibliography{refs.bib} 

\end{document}